\documentclass[prl,superscriptaddress,notitlepage,twocolumn,nofootinbib]{revtex4-1}

\usepackage{amsmath}    
\usepackage{graphicx}   
\usepackage{verbatim}   
\usepackage{color}      
\usepackage{hyperref}   
\usepackage{amsfonts}
\usepackage[english]{babel}
\usepackage{ucs}
\usepackage[utf8x]{inputenc}
\usepackage{subfigure}

\newcommand{\Corr}{\operatorname{Corr}}

\begin{document}

\title{Random matrix analysis of Ca$^{2+}$ signals in $\beta$-cell collectives}

\author{Dean Korošak}
\affiliation{University of Maribor, Faculty of Medicine, Institute for Physiology, Maribor, Slovenia}
\affiliation{University of Maribor, Faculty of Civil Engineering, Transportation Engineering and Architecture, Maribor, Slovenia}

\author{Marjan Slak Rupnik}
\thanks{Corresponding author:\\marjan.slakrupnik@meduniwien.ac.at}
\affiliation{Medical University of Vienna, Center for physiology and pharmacology, Vienna, Austria}
\affiliation{University of Maribor, Faculty of Medicine, Institute for Physiology, Maribor, Slovenia}
\affiliation{Alma Mater Europaea - European Center Maribor, Maribor, Slovenia}

%\end{comment}

\date{\today}

\begin{abstract}

Even within small organs like pancreatic islets, different endocrine cell types and subtypes form a heterogeneous collective to sense the chemical composition of the extracellular solution and compute an adequate hormonal output. Erroneous cellular processing and hormonal output due to challenged heterogeneity result in various disorders with diabetes mellitus as a flagship metabolic disease. Here we attempt to address the aforementioned functional heterogeneity with comparing pairwise cell-cell cross-correlations obtained from simultaneous measurements of cytosolic calcium responses in hundreds of islet cells in an optical plane to statistical properties of correlations predicted by the random matrix theory (RMT). We find that the bulk of the empirical eigenvalue spectrum is almost completely described by RMT prediction, however, the deviating eigenvalues that exist below and above RMT spectral edges suggest that there are local and extended modes driving the correlations.
We show that empirical nearest neighbor spacing of eigenvalues follows universal RMT properties regardless of glucose stimulation, but that number
variance displays clear separation from RMT prediction and can differentiate between empirical spectra obtained under non-stimulated and stimulated
conditions. We suggest that RMT approach provides a sensitive tool to assess the functional cell heterogeneity and its effects on the spatio-temporal dynamics a collective of beta cells in pancreatic islets in physiological resting and stimulatory conditions.

\end{abstract}

\maketitle 

\section{Introduction}

Pancreatic islets are collectives of endocrine cells. Based on the end hormone that these cells exocytose in a Ca$^{2+}$-dependent manner after being stimulated, several types of cells have been described to compose an islet, with 3 dominant cell types: alpha, beta and delta~\cite{briant2017functional}. Islets from different parts of pancreas can contain different fractions of each of these cell types, but the bulk cellular mass in a typical islet is in an non-diabetic organism composed of a collectives of insulin secreting beta cells~\cite{dolenvsek2015structural,rorsman2017pancreatic}. Early studies assumed these beta cell collectives to be a rather homogeneous population of cells, however, the subsequent functional analyses have revealed a remarkable degree of heterogeneity even in dissociated beta cells in culture. The beta cells were found to differ in a number of physiological parameters, among others in glucose sensitivity and Ca$^{2+}$ oscillation pattern~\cite{zhang2003ca2+}, electrical properties~\cite{misler1986metabolite}, redox states~\cite{kiekens1992differences}, or pattern on cAMP oscillations~\cite{dyachok2006oscillations}. These early quests~\cite{pipeleers1992heterogeneity} have been mostly searching for morphological, physiological and molecular features that would presumably satisfy at least 3 criteria: a) entitle special roles for individual cells within the collectives, b) remain valid even after cell dissociation, and c) enable to trace embryonic and postnatal development as well as changes during pathogeneses of different forms of diabetes. Recent onset of efficient high-throughput analyses has catapulted these approaches on mostly dissociated cells to a completely new level and enabled identification of a multitude of functional and non-functional subpopulations with their functional characteristics, gene and protein expression, incidences and diabetes-related changes (for a recent review refer to~\cite{benninger2018new}). One of the main results of these analyses is that the subpopulations described in different studies have relatively little in common and currently their translational relevance is limited. This is, however, not surprising since these approaches primarily deal with sample averages and present merely a small number of discrete snapshots of a very dynamic complex activity. Nevertheless, they turned out to be extremely useful in an attempt to construct a pseudotime map of the sequence of events in pancreatic endocrine system~\cite{damond2019map} and its interaction with the immune system~\cite{wang2019multiplexed} during the progression of type 1 diabetes mellitus (T1D) in humans.

The fact is that most of what we know about pancreatic beta cells has been gained by studying dissociated beta cells in cell culture. Therefore, even phenomena that can only be observed in isolated groups of electrically coupled beta cells, like electrical activity~\cite{rorsman1986calcium} or cytosolic Ca$^{2+}$ oscillation,  are currently still mostly modelled within the framework of a single cell excitability~\cite{sherman1988emergence,bertram2018closing}. However, each beta cell interacts with several immediate and distal neighboring cells in a pancreatic islet, implicating high-ordered interactions between a large number of elements. Therefore, there is a rich exchange of signals within such a beta cell collective, both through direct cell-cell coupling~\cite{bavamian2007islet} as well as through paracrine signalling~\cite{caicedo2013paracrine,capozzi2019beta}. Such an organisation necessarily yields a complex response patterns of cell activity after stimulation with physiological or pharmacological stimuli. Until recently, the richness of cell-cell interactions also could not be experimentally documented. However, recent technological advancements made it possible to use various optical tools to address these issues~\cite{frank2018optical}. For example, the functional multicellular imaging (fMCI) enabled completely new insights into our understanding of a beta cell in an islet as a biological network~\cite{stovzer2013glucose,stovzer2013functional,dolenvsek2013relationship}. The dynamics of a measurable physiological parameter can namely be recorded in hundreds of beta cells within their intact environment~\cite{speier2003novel,marciniak2014using} simultaneously. The measured oscillatory cytosolic Ca$^{2+}$ concentration changes, which are required to drive insulin release turned out to be a practical tool to trace cellular activity and fundamental to study their interactions in such big collectives ~\cite{dolenvsek2013relationship,stovzer2013glucose,stovzer2013functional}. With the use the tools of statistical physics we constructed, for example the network topology in beta cell activation, activity and deactivation during transient glucose challenges~\cite{stovzer2013functional,markovivc2015progressive, johnston2016beta,gosak2015multilayer}. As in some other, previously analysed biological systems, also for the pancreatic islets, the minimal model incorporating pairwise interactions provides accurate predictions about the collective effects~\cite{schneidman2006weak,korovsak2018collective}. 

Along these lines we have recently shown that beta cell collectives work as a broad-scale complex networks~\cite{stovzer2013functional,markovivc2015progressive,gosak2017network}, sharing similarities in global statistical features and structural design principles with internet and social networks~\cite{barabasi2016network,daniels2016quantifying,duh2018collective, milo2002network,perc2017statistical}. In addition to complex network description when strong cell-cell interaction are primarily taken into account, the analyses of weak pairwise interaction enabled us to use a spin glass model~\cite{korovsak2018collective}, as well as the assessment of self organized criticality~\cite{bak2013nature,markovic2014power,gosak2017critical}, (ref Sto\v zer this issue) also often found in biological samples~\cite{schneidman2006weak}. The important result from these functional studies is that a faulty pattern of hormone release due to deviating numbers of individual cell types or changes in their function lead to one of the forms of a large family of metabolic diseases called diabetes mellitus~\cite{american2014diagnosis,nasteska2018role,pipeleers2017heterogeneity,skelin2017triggering,capozzi2019beta}.

%%% basic RMT
The basic object that we study here is the correlation matrix $\mathbf{C}$ with elements computed from measured Ca$^{2+}$ signals:
\begin{equation}
     \Corr(y_i,y_j) = C_{ij} = \frac{\langle y_i y_j\rangle - \langle y_i\rangle \langle y_j\rangle}{\sigma_i\sigma_j},
\end{equation}
where $y_i(t)$ is the i-th time series of Ca$^{2+}$ signal out of $N$ signals measured simultaneously in a collective of pancreatic beta cells.

Random matrix theory~\cite{mehta2004random,guhr1998random} (RMT) is concerned with statistical properties of matrices with random elements. Applying RMT to 
correlation matrices, we study the spectrum of the correlation matrix $\mathbf{C}$ given by the set of its eigenvalues $\lambda_n$:
\begin{equation}
    \mathbf{C}\mathbf{u}_n=\lambda_n \mathbf{u}_n,
\end{equation}
where $\mathbf{u}_n$ are the corresponding eigenvectors. 

Statistical properties of the spectra of random correlation matrices for $N$ uncorrelated time series with $M$ random elements 
where $q = N/M$ is finite in the limit $N,M \to\infty$ are known analytically~\cite{marchenko1967distribution, bun2017cleaning}. 
The eigenvalue probability density is: 

\begin{equation}
    \rho(\lambda) = \frac{1}{2\pi q\lambda}\sqrt{(\lambda_+ - \lambda)(\lambda - \lambda_- )},
\end{equation}

where the spectral boundaries are:

\begin{equation}
    \lambda_\pm = \left(\sqrt{q}\pm 1\right)^2
\end{equation}

When the spectrum of the correlation matrix is unfolded~\cite{guhr1998random} by mapping eigenvalues $\lambda_k \to \xi_k$ so that the 
probability density of the unfolded eigenvalues is constant $\rho(\xi) = 1$, the RMT predicts that the distribution $P(s)$ of nearest neighbor
spacings $s_k = \xi_{k+1}-\xi_k$ is approximately given by the Wigner surmise~\cite{mehta2004random}:
\begin{equation}
    P(s) = \frac{\pi}{2}s\exp{\left(-\frac{\pi}{4}s^2\right)}.
\end{equation}

Possible pair correlations in the eigenvalue spectrum on scales larger than nearest neighbors can be revealed with the use of variance of $n_{\xi}(L)$, the number of 
eigenvalues in the interval of length $L$ around eigenvalue $\xi$. This number variance~\cite{mehta2004random} is given by:
\begin{equation}
    \Sigma^2(L) = \langle \left(n_{\xi}(L) - L\right)^2\rangle .
\end{equation}
If the eigenvalue spectrum is poissonian the number variance is $\Sigma^2(L)\sim L$, while
real, symmetric random matrices exhibit correlated spectra for which RMT predicts $\Sigma^2(L)\sim \log L$~\cite{mehta2004random}.

Previous work using RMT in different systems, e.g. on statistics of energy levels of complex quantum systems~\cite{guhr1998random,mehta2004random} or correlations in financial markets~\cite{plerou2002random} identified that a bulk of the eigenvalue spectrum agreed with RMT predictions, which suggested a large degree of randomness in the measured cross-correlations in these systems. Only a small fraction of typically a few percent of eigenvalues were found to deviate from universal RMT predictions and were instrumental to identify system specific, non-random properties of the observed systems and yielding key information about the underlying interactions. 
%In addition, universal properties like eigenvalue number spacing and number variance of the correlation from the financial market were completely predicted described by %GOE statistics, suggesting strongly correlated eigenvalues fluctuations. 
In biological systems, RMT has been used to filter out critical correlations from data-rich samples in genomic, proteomic and metabolomic analyses~\cite{luo2006application,agrawal2014quantifying}, as well as in brain activity measured by EEG~\cite{vseba2003random}. 
%Determining the Brody parameter for the nearest neighbor spacing distribution in these biological systems peaks at the value of 1, suggesting GOE distribution, which has %been interpreted as the existence of minimal amount of randomness. 
While eigenvalue spacing distributions showed agreement with RMT predictions, the number variance distributions often display deviations pointing to the physiologically relevant reduction in correlated eigenvalues fluctuations with partially decoupled components transiting towards Poisson distribution~\cite{vseba2003random}. Such transitions have also been used as an objective approach for the identification of functional modules within large complex biological networks~\cite{luo2006application}. Additionally, as for protein-protein interactions in different species, these latter deviation from RMT predictions has been interpreted as an evidence to support the prevalence of non-random mutations in biological systems~\cite{agrawal2014quantifying}. 

In this paper we used the RMT approach to test the cross-correlations in the cytosolic Ca$^{2+}$ oscillations under non-stimulated and glucose stimulated conditions. We demonstrate that statistical properties of cross-correlations based on functional multicellular imaging data follows those predicted by RMT, with both high- and low-end deviating eigenvalues, suggesting local as well as global modes driving this correlation in functional islet. In addition, our results show that the long range correlations in eigenvalue spectrum deviate in a stimulus dependent manner.

\section{Dataset description}

We define beta cell collective activity to sense nutrients and produce metabolic code as the relevant constraining context for the physical outcomes of analysis (\cite{ellis2018dynamical,korovsak2018collective}). Our data consist of Ca$^{2+}$ activity recorded by multicellular imaging on islets of Langerhans from fresh pancreatic slices. We used raw data for each calcium signal, but we detrended the signals to remove possible sources of spurious correlations due to systematic slow variations caused by the imaging technique. A common problem in the analysis of fMCI Ca$^{2+}$ signals in living tissue is selection of regions of interest corresponding to a true signal originating from a cytosolic area of an individual cell and not two or more neighbouring cells. In practice the reproducibility of the results depends on the level of experience of the operator to subjectively recognize structure from the patterns of activity. While we are primarily interested in the activity of a large population of cells, their interactions/correlations and their collective response, it is crucial that this signals originate from regions of interest that correspond to individual cells. Collectives of cells, like beta cells in the islet of Langerhans are densely packed structures, where extracellular space and the cell membrane represent a relatively small to negligible cross-section area on the image of two-dimensional optical section obtained by confocal microscopy. Therefore we decided to avoid the aforementioned subjectivity issue by the random sampling of pixel level signals in the recorded time series. For this analysis we randomly selected $N = 4000$ signals out of the complete dataset of 256x256 signals each $M = 23820$ timesteps long (about 40 minute recordings at 10 Hz resolution). Glucose concentration was changed during the recording from 6 mM (lasting about first 5000 timesteps) to 8 mM and back to 6 mM (approx. last 5000 timesteps) near the end of experiment.  

%In this study, we tested RMT approach in a attempt to better understand the structure of our data and the robustness of the RMT measures we can extract from both %non-stimulated and stimulated conditions. To quantify correlations we calculate Ca2+ concentration change in a randomly selected pixel within a cell i=1,...N over time %scale delta t

%where xi(t) denotes the Ca2+ value of a pixel. We did not correct for variability of standard deviations between the pixels. We then computed the equal-time %cross-correlation matrix C with elements:

%The elements Cij are restricted to a domain -1 < Cij < 1, which describes a domain between perfect correlations (Cij=1) and perfect anti-correlations (Cij=-1). Cij=0 is %valid for uncorrelated events. We assumed that the cross correlation between the pixels to be stationary in the observed time frame and we are aware of limitations of a %finite length of the time series available. We used Cij values to identify which cell-cell interactions remained correlated in the recorded period and we tested its %statistics against a random correlation matrix. In the event of a complete conformity between both matrices also the content of the empirical matrix is random. %Alternatively, deviations in structure of matrix C could contain information about existing correlation rules within the measured matrix.

The source of correlations in a cell population where the terminal action is a calcium-dependent process (e.g. exocytosis of insulin in beta cells) are the individual events in a form of plasma membrane ion channel or transporter activity, internal membrane ion channel or transporter activity, as well as calcium leak from activated immediate neighboring beta cell (\cite{berridge2000versatility}). The correlations between the activities of beta cells depend strongly upon the glucose concentration(\cite{markovivc2015progressive,dolenvsek2013relationship}), however in the physiological plasma glucose range (6-9 mM), most correlations are weak(\cite{korovsak2018collective}), so that the probability of detecting co-activation basically equals the product of the probabilities of activities of individual beta cells. The correlations are statistically significant for almost all pairs of immediate neighbors.

\section{Results}

\begin{figure*}[t]
\centering
\includegraphics[scale=0.5]{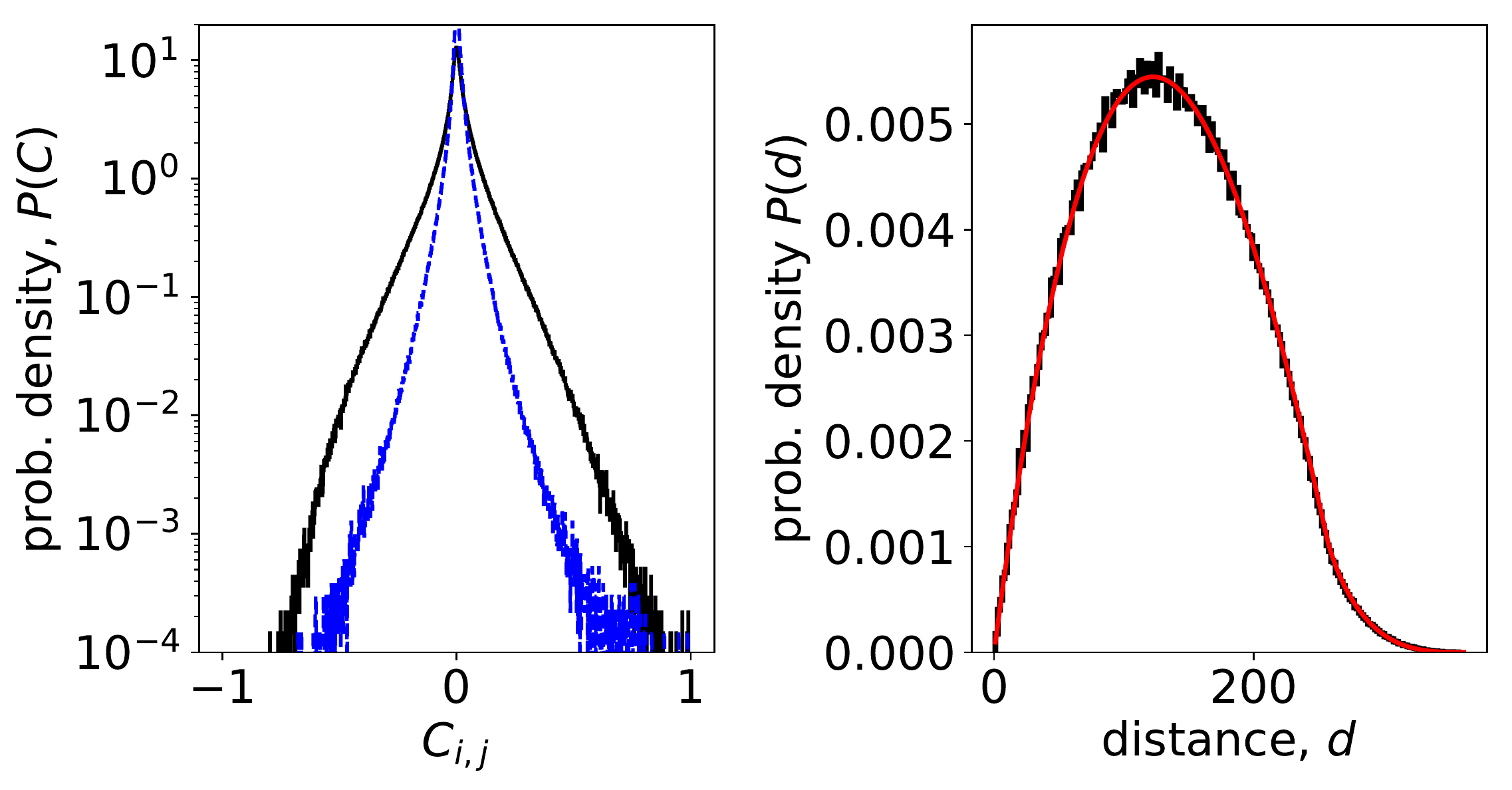}
\caption{\label{fig:fig1} (left) Correlation coefficient distribution of $N=4000$ Ca$^{2+}$ signals randomly chosen from experimental dataset (outer, black line). Dashed blue line is the distribution of correlations between residuals after removing the influence of the largest eigenvalue in signals. (right) Distribution of distances between signals randomly chosen from Ca$^{2+}$ imaging data. Black line is the experimental data, thinner red line is the theoretical distribution of distances between random points in a square~\cite{philip2007probability}.}
\end{figure*}

%First, we computed the correlation matrix \mathbf{C} containing all correlation coefficients between $N=4000$ Ca$^{2+}$ signals in the dataset. 

The distribution of correlation coefficients reveals that most of the correlations are weak, but there is also non-negligible contribution of highly correlated pairs of signals (Figure 1, left, black outer line). We also checked the sampling procedure by comparing the computed distribution of distances between pairs of randomly chosen points from 256x256 image square to the analytical probability distribution of distances between two random points in a square~\cite{philip2007probability} (Figure 1, right). We found a perfect match between the distance distribution computed from data and the theoretical distribution, confirming that our random sampling of data points was non-biased.

Guided by the observed non-gaussian nature of correlation distribution (Figure 1, left) we explored a detailed structure of the correlation matrix, since distribution of correlation coefficients only partially hints to the nature of cell to cell coordination. 
To this end we computed the eigenvalues and eigenvectors of the correlation matrix (Eq. 2) and compared the obtained eigenspectrum with the RMT prediction. 
In Figure 2 (top left) we show the distribution of eigenvalues that belong to the empirical correlation matrix (black trace) and the RMT prediction 
(red line) given by Eq. 3. While most of the eigenvalues falls within the limits $\lambda_\pm$ of the RMT spectrum, there are also significant deviations from 
RMT prediction. We found the largest empirical eigenvalue $\lambda_{max}$ two orders of magnitude away from the upper limit of the RMT spectrum, and also a part of the 
empirical spectrum that extends below the lower RMT limit. To see if the deviations from the RMT are inherent to the measured Ca$^{2+}$ signals, we 
prepared a surrogate dataset by randomly shuffling each signal's time series. We then computed the correlation matrix and its eigenvalue spectrum from randomized surrogate dataset. As shown in Figure 2 (top right), the match between the eigenvalue distribution of randomized dataset and RMT is perfect.   

%% largest eigenvalue market mode
Previous RMT analysis of stock correlations in financial markets consistently showed~\cite{laloux1999noise,plerou1999universal,plerou2002random} that the distribution of components of the eigenvector $\mathbf{u}_{max}$ corresponding to largest eigenvalue $\lambda_{max}$ strongly deviates from Gaussian form, suggesting that this mode reflects the collective response of the system to the stimuli. In our case this corresponds to collective response of beta-cells to glucose stimulus. In a linear statistical model for Ca$^{2+}$ signals, we model the response common to all beta-cell with $Y(t)$ and the signals are expressed as:
\begin{equation}
    y_i(t) = a_i + b_i Y(t) + \delta y_i(t),
\end{equation}
where $\delta y_i(t)$ is the residual part of each signal. Coefficients $a_i,  b_i$ are obtained by regression. Following~\cite{plerou2002random} we 
approximated the common response $Y(t)$ with the projection of all signals on the largest eigenvector:
\begin{equation}
    y_{max}(t) = \sum_{i=1}^{N}u_i(\lambda_{max})y_{i}(t),
\end{equation}
where $u_{max,i}$ is the i-th component of the eigenvector corresponding to largest eigenvalue $\lambda_{max}$. To see the influence of the collective response 
to the distribution of correlation coefficents, we computed using $Y = y_{max}$ the residuals $\delta y_i(t)$ for all $N$ signals and their correlations 
$C_{res(i,j)} = \Corr(\delta y_i,\delta y_j)$. The dashed blue line (inner trace) on Figure 1 (left) shown the distribution of $\mathbf{C}_{res}$ and reveals 
that the collective response predominantly contributes to large correlations. 

\begin{figure*}[t]
\subfigure{\includegraphics[scale=.5]{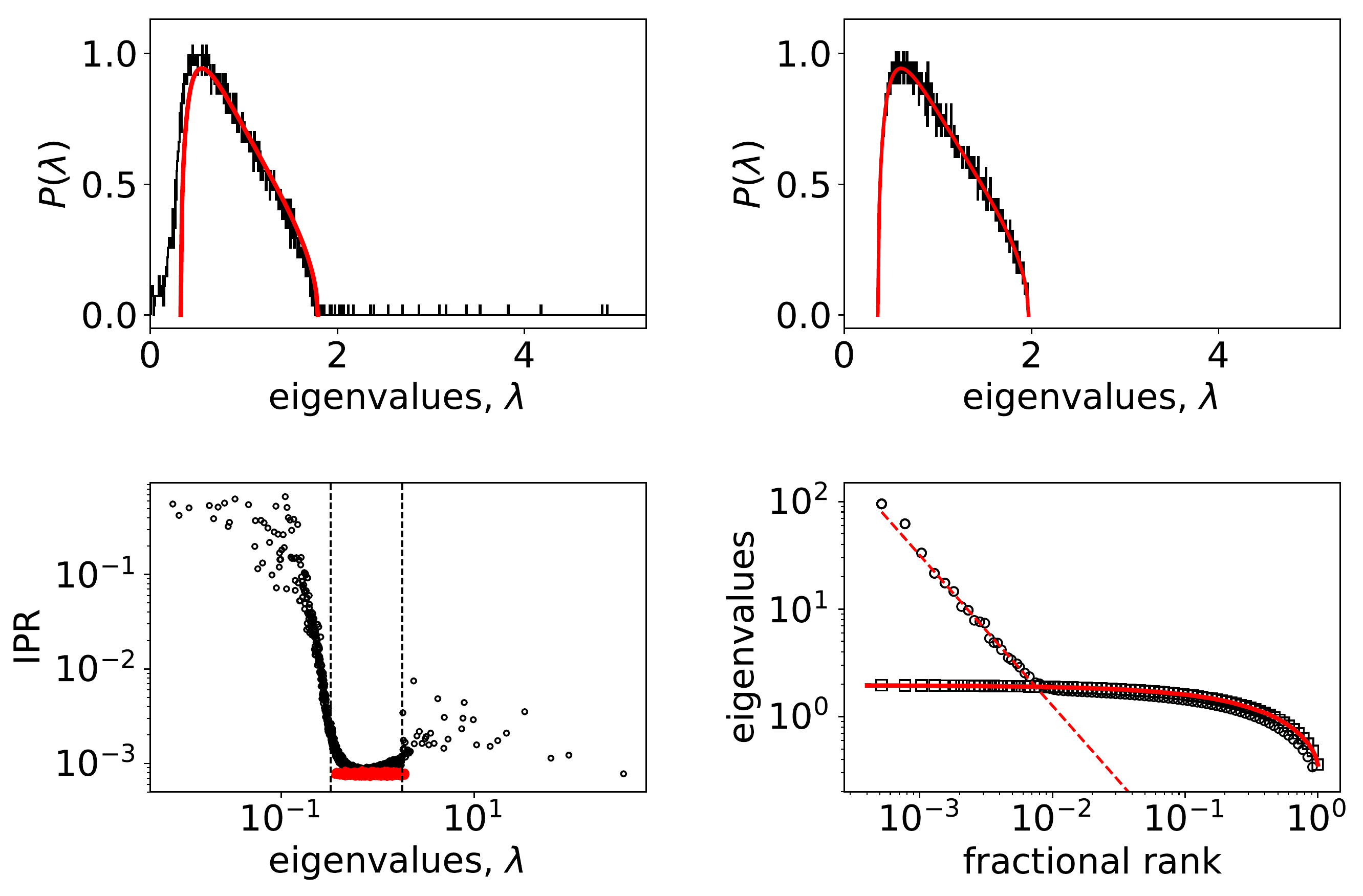}}
\caption{ \label{fig:fig2} (top left) Probability distribution of eigenvalues of the empirical correlation matrix for $N=4000$ randomly picked signals (black solid line) compared to the distribution of eigenvalues of random correlation matrix  of the same size (red solid line). (top right) Probability distribution of eigenvalues of the surrogate correlation matrix constructed from shuffled empirical values (black solid line) compared to the random correlation matrix of the same size (red solid line). (bottom left) Inverse participation ratio plot of eigenvalues showing a random band matrix structure of C with large IPR values at both edges of the eigenvalue spectrum. The dashed vertical lines show RMT bounds. The IPR spectrum for randomized correlations is shown in red. (bottom right) The fractional rank plot of the entire spectrum of eigenvalues (open dots). For comparison we added the same plot of eigenvalues of correlation matrix computed from randomized data (open squares). The full line shows 
the fractional rank plot of eigenvalue spectra obtained from distribution given by eq. 3. The shape of the distribution of large eigenvalues points to a scaling relationship.}
%\textbf{}
\end{figure*}

To test further if the largest eigenvalue and the corresponding eigenvector capture the collective calcium response we compared the average signal 
$\overline{y}(t) = 1/N\sum_{j}y_j(t)$ with $y_{max}$. The correlation between signals projected on the largest eigenvalue mode and mean signal was high: $Corr(y_{max},\overline{y})\approx 0.8$, confirming the expectation that the largest eigenvalue represents collective effect. Similarly, we checked
how similar are the signals corresponding to the bulk RMT eigenvalues:

\begin{equation}
    y_{bulk,i}(t) = \sum_{j=1}^{N}u_j(\lambda_i)y_{j}(t),
\end{equation}
where $\lambda_i$ is the eigenvalues from the RMT interval $[\lambda_+,\lambda_-]$. The computed correlation between signals projected on 
bulk eigenvectors and mean signal, averaged over all signals was $<\Corr(y_{bulk,i}, \overline{y})> = -0.0044\pm 0.0047$ suggesting no correlation between the mean signal and signals coming from the bulk RMT regime. 
%%% ipr
To further characterize the eigenvector structure of the empirical Ca$^{2+}$ correlation matrix, we looked at the inverse participation ratio (IPR) of eigenvector
$\mathbf{u}(\lambda)$ corresponding to eigenvalue $\lambda$ defined as~\cite{plerou1999universal,plerou2002random}: 
\begin{equation}
    I(\lambda) = \sum_{j}^{N}\left(u_{j}(\lambda)\right)^4.
\end{equation}
The value of $1/I(\lambda)$ reflects the number of nonzero eigenvector components: if an eigenvector consist of equal components $u(\lambda)_i = 1/\sqrt(N)$ then 
$1/I(\lambda) = N$, in other extreme case $1/I(\lambda) = 1$ when an eigenvector has one component equal to 1 and all others are zero. 

Figure 2 (lower left) shows the computed values of IPR for all eigenvectors as function of corresponding eigenvalue. The red datapoints are the IPR data computed 
for the surrogate, randomized timeseries data for which we found $1/I \sim N$ as expected. We found similarly values $1/I \sim N$ for the largest eigenvalues of the empirical spectrum (black datapoints) suggesting that to this eigenvectors almost all signals contribute. Deviations from flat RMT prediction at the edges of the RMT spectrum ($[\lambda_+,\lambda_-]$ interval, vertical dashed lines) with large $I(\lambda)$ values suggests that these states are localized with only a few signals contributing. This points to a complex structure of the empirical correlation matrix $\mathbf{C}$ with coexisting extended and localized eigenvectors similar to 
one found in correlations in financial markets~\cite{plerou1999universal,plerou2002random}. 
%fractional rank
In addition, as shown in Figure 2 (lower right, open dots), we observe a scaling behavior in rank-ordered plot of 
eigenvalues of empirical correlation matrix that has been connected with a fixed point in renormalization group sense~\cite{bradde2017pca,meshulam2018coarse}. For 
comparison, we plot also the rank-ordered eigenvalues of randomized data (open squares) and RMT prediction based on eigenvalue density given by Eq. 3 (full line) 
which perfectly describes the randomized dataset. The observed scaling of eigenvalues hints towards the critical behavior that was conjectured for 
beta-cell collective at the transition from glucose non-stimulated to stimulated phase (from 6 mM to 8 mM)~\cite{gosak2017critical}.  

To explore the statistical differences of signals in non-stimulated and stimulated phase, we separated the original data into two groups of $N$ signals each with $M=10^4$ timesteps corresponding to response to 6 mM and 8 mM glucose stimuli. For each group we computed the unfolded eigenvalue spectra and also for randomized data. The results for the nearest-neighbor spacing and number variances are shown in Figure 3. For nearest-neighbor spacing distribution we find a good agreement with the RMT prediction both, for non-stimulatory and stimulatory conditions, as well as shuffled stimulated data. All three datasets are well described with the Wigner surmise (Eq. 5), so nearest-neighbor spacing does not seem to be sensitive to stimuli changes. On the other hand, however, the number variance is sensitive to stimuli change already during physiological stimulation of the beta-cell collective. The random matrix prediction is in this case valid for shuffled stimulated data only (Figure 3, right). 

\begin{figure*}[t]
\centering
\includegraphics[scale=0.5]{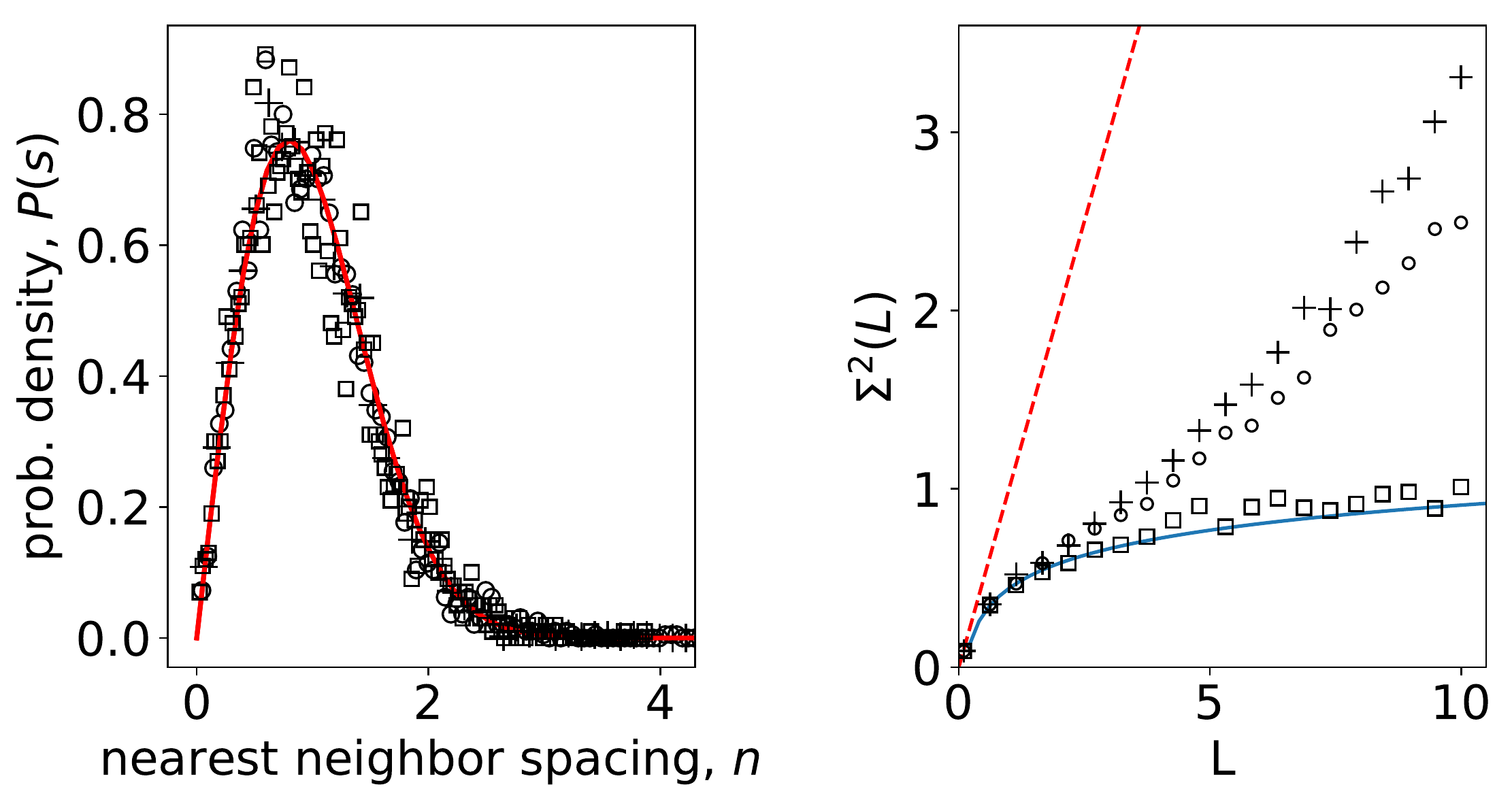}
\caption{\label{fig:fig3} (left) Nearest-neighbor spacing distribution of empirical correlation matrix eigenvalues of calcium signals in non-stimulatory and 
stimulatory regime. Open squares: shuffled, randomized data; open dots: 8 mM glucose; crosses 6 mM glucose; full line Wigner surmise (eq. 5). (right) Number variance of eigenvalue spectra of calcium signals. Open squares: shuffled, randomized data; open dots: 8 mM glucose; crosses 6 mM glucose; full line RMT prediction $\Sigma^2(L) = 1/\pi^2 \left( \log(2\pi L) + 1+\gamma - \pi^2/8 \right)$~\cite{mehta2004random}; dashed line Poissonian limit $\Sigma^2(L) = L$}. 
\end{figure*}

\section{Discussion}

The unique spatio-temporal resolution of functional multicellular imaging and sensitivity of advanced statistical approaches for a plethora of different modes of complex network scenarios and levels of criticality, makes these approaches a method of choice to assess the nature of cell-cell interactions under different stimulation conditions. At the same time it enables us to test the validity of experimental designs for study of beta cell function, primarily in the domain of stimulation strength and dynamics. We suggest that without such validation the most critical events in the activation chain within the beta cell collective have been and shall be overlooked or misinterpreted (Stožer et al., current issue). The predominant use of supraphysiological glucose concentration can namely severely deform the relatively slow beta cell recruitment in a collective at physiological glucose concentrations~\cite{gosak2017network,stovzer2013glucose} and miss the typically segregated network clusters of Ca$^{2+}$ events~\cite{markovivc2015progressive,benninger2014cellular,westacott2017spatially}, turning critical behavior into disruptive supracritical activity~\cite{gosak2017critical}. Only under rather narrow physiological conditions it shall be possible to extract the fine structure of cell-cell interactions causing long-term and efficient cell collaboration with the collective. Breaking apart this delicate structure of cell-cell interaction does result in a massive activity, which can be readily described by tools of physics, but this activity does not necessarily serve its physiological or biological purpose~\cite{ellis2018dynamical}.     

A common denominator of the previous attempts to categorize different beta cell types points to some metabolic and secretory features that can be either reproduced between different classifications or not. Usually there exist a bulk of one subtype and one or more less frequent subtypes~\cite{benninger2018new}. These less frequent subtypes can nevertheless have important regulatory roles that may not be immediately apparent. This issue is particularly critical if the frequency of a beta cell subtype represents only a couple of percent of the entire beta cell population in an islet. Along these lines there have been some indications regarding the beta cell subtypes that can serve as pacemakers or hubs within a dynamic islet cell network~\cite{johnston2016beta,lei2018beta}, however due to the nature of complexity of network features, we may still be short of evidence for definitive conclusions. The full description of heterogeneity of endocrine cells within an islet, ultimately producing a adequate release of hormones is therefore still lacking. In trying to grasp this complexity, it is important to take into account interaction of beta cell collectives with other cell types in and around an islet, like glucagon-secreting alpha cells~\cite{svendsen2018insulin,capozzi2019beta} or somatostatin secreting delta cells~\cite{rorsman2018somatostatin} as well as neurons and glial cells~\cite{meneghel2004vivo}, but also endothelial, immune cells~\cite{damond2019map}, as well as acinar and ductal cells~\cite{bertelli2005association}. 

Random matrix theory is a fitting mathematical framework which provides powerful analytical tools to separate cell-cell interactions happening by chance from those produced by specific coordinated interactions after a changed chemical composition of cell´s surrounding. In the financial sector, adequate asset allocation and portfolio risk-estimation can lead to a higher profit and is therefore clear why it makes sense to invest time into cross-correlation analyses~\cite{plerou2002random}. But what would be the gain of knowing that randomness of cell-cell correlation matrices is physiologically regulated? Firstly, we suggest that the analysis of the universal properties of empirical cross-correlations is a valuable tool to identify distinct types and further subtypes of endocrine cells within an islet through their non-local and local effects. The largest eigenvalue of $\mathbf{C}$ namely represents the influence of non-local modes common to all measured Ca$^{2+}$ fluctuations. Other large eigenvalues can be used to address cross-correlations between cells of the same type, cells with specific functions in the collective or that these cells reside in topologically similar area of the islet. Quantifying correlations between different beta cells in an islet is therefore an exciting scientific effort that can help us understand cell communities as a complex dynamical system. Our results show that the number variance reflecting the correlation between $L$ subsequent eigenvalues (a measure for long range correlations in eigenvalue 
spectrum) follows the RMT predictions up to a certain distance $L$, however at larger distances it starts to deviate in a stimulus dependent manner, suggesting structural features in the beta cell network. Transitions between Poissonian and symmetric matrices in biological systems have been previously described during the process of either integration or segregation of complex biological networks, showing various degrees of long range correlations at various physiological conditions~\cite{luo2006application}. This understanding has a vital practical value since it can help decipher different roles that beta cells can play in a collective and to further validate the importance, if any, of previously defined and continuously appearing novel molecular markers of beta cell heterogeneity~\cite{benninger2018new,damond2019map,wang2019multiplexed}. An advanced knowledge about the dynamic properties of the functional cell types will shed a new light into understanding of physiological regulation of insulin release and the assessment of perils of stimulation outside of the physiological range. Furthermore, it can help us elucidate the mechanisms on how this function changes during the pathogenesis of different forms of diabetes mellitus and lead us to novel approaches of therapy planning and prevention.

\subsection*{Acknowledgements}

The authors DK and MSR, acknowledge the financial support from the Austrian Science Fund/Fonds zur Förderung der Wissenschaftlichen Forschung [Bilateral grant I3562-B27 (to M.S.R.)] and the Slovenian Research Agency (research core funding, No. P3-0396), as well as research project, No. N3-0048).

\bibliography{refs}{}
\bibliographystyle{apsrev4-1}

\begin{comment}
%%%% discuss RG and PCA approach (bialek)

correlation between signals projected on the largest eigenvalue mode and mean signal $Corr(max,mean)\approx 0.8$, correlation 
between signals projected in bulk eigenvalues and mean signal (averaged over RMT part of the spectrum) $<\Corr(bulk, mean)> = -0.0044\pm 0.0047$

\begin{equation}
    P_{ij} = \sum_{\lambda}u_i(\lambda) u_j(\lambda)
\end{equation}

in RMT we expect that variance diminishes as we integrate out the lower eigenvalue modes

mean signal or projection to largest eigenvalue 2.07, 2.68

projection onto bulk = gaussian $3.062 \pm 0.001$
\begin{equation}
    \psi_i = \sum_{j}P_{ij}y_j
\end{equation}

all eigenvalues signal variance: $13.7 \pm 6.9$ 

separated nonstimulated and stimulated parts:
variance of all eigenvalue signals at 8 mM, 6 mM, and shuffled signals: 7.9, 3.99, 4.16
\end{comment}

\end{document}